\documentclass[fleqn,useAMS,usenatbib]{mnras}
\usepackage{amsmath,subfig,footmisc,float}
\usepackage{graphicx,txfonts,fleqn,enumerate,color,hyperref,listings,booktabs,longtable}
\usepackage{ulem} % for sout

\renewcommand*{\v}[1]  {\boldsymbol{#1}}

\newcommand   {\sub}[2]  {{#1}_{\mathrm{#2}}}

\newcommand*  {\twovector}[2] {{\begin{pmatrix} $1 \\ $2 \end{pmatrix}}}

\renewcommand {\emph}[1]  {\textit{#1}}

\title[{Time-reversible timestepping}]
{\boldmath Time-reversible timestepping in many-body planetary systems}

\author[Hernandez et al.]
	{David M.~ Hernandez\href{http://orcid.org/0000-0001-7648-0926}{\includegraphics[width=11pt]{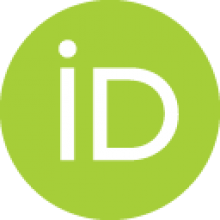}}$^{1,2}$\thanks{Email: dmhernandez@ntnu.edu.tw},
    Tenri Jinno\href{https://orcid.org/0009-0001-5384-4202}{\includegraphics[width=11pt]{orcid-ID}}$^{3}$, and 
    Junichiro Makino$^{3}$\\ 
	$^1$ Department of Physics, National Taiwan Normal University, Taipei 116, Taiwan\\
	$^2$ Department of Astronomy, Yale University, Kline Biology Tower, 219 Prospect St., New Haven, CT 06511, USA \\
	$^3$ Department of Planetology, Graduate School of Science, Kobe University, 1-1, Rokkodai-cho, Nada-ku, Kobe 657-8051, Japan \\
	}
\begin{document}

\maketitle

\label{first page}
%%%%%%%%%%%%%%%%%%%%%%%%%%%%%%%%%%%%%%%%%%%%%%%%%%%%%%%%%%%%
\begin{abstract}
Simulating chaotic planetary system evolution requires resolving small time-scales due to close encounters, scattering, collisions, and more.  Simulations typically assign fixed timesteps to all particles, however, which is inefficient and wastes compute resources in a chaotically evolving system; moreover, it may under-resolve the simulation.  Hybrid methods are a blunt tool for resolving small time-scales as they do not allow for fine-grained adaptation.  Assigning individual timesteps, which vary in time, to planetary bodies usually violates desirable properties such as symplecticity and reversibility.  In this work, we develop an ``almost'' exactly time-reversible direct $N$-body integrator, \texttt{MTR}, in which each pair of planets gets its own timestep, which itself can vary as the system evolves.  In contrast to our previous work, we achieve this adaptation on the fly, independent of a global timestep.  We provide a \texttt{C} version of the code and test it on several problems, demonstrating no energy error drift, as expected of a reversible method.  At large particle numbers, up to $N = 975$, we compare against the state-of-the-art \texttt{GPLUM} code, which we can use in direct Hermite mode (for which it is not optimized) or tree mode (for which it is).  \texttt{MTR} provides a speedup of up to two orders of magnitude compared to the Hermite mode, while agreeing with the distribution of orbits.  In our default configuration, \texttt{MTR}'s speed matches or improves on the tree mode's speed, by up to an order magnitude, despite the tree mode's faster and approximate force calculation.  
\end{abstract}
%%%%%%%%%%%%%%%%%%%%%%%%%%%%%%%%%%%%%%%%%%%%%%%%%%%%%%%%%%%%
\section{Introduction}
\label{sec:intro}
One of the challenges of simulating gravitational systems \citep{aar08,DR11,HH03} is the lack of a fixed length-  and time-scale over which gravity acts.  A planetary system has longer secular time-scales ($\approx 10^5$ yr for the Solar system), orbital time-scales ($\approx 100$ yr for the Solar system), and close-encounter and scattering time-scales, which can be arbitrarily small.  There are different strategies for solving planetary systems.  Secular codes \citep{Laskar1985,Laskar1988,Laskar1990,HaddenTamayo2022} are designed to study long-term, slowly varying dynamics and can take large timesteps, but will miss any features occurring on shorter scales.  Symplectic methods like the Wisdom--Holman method \citep{WH91,Kinoshitaetal91} resolve orbital time-scales, but the timestep is limited by the shortest orbit of the system.  This is inefficient.  Additionally, this orbital time-scale itself can vary in time, making the chosen timestep useless.  The obvious solution is just to change the timestep, but this typically ruins several beneficial properties of the method \citep{SkeelGear1992,cal93,SB94,leim04,hair06,blanescasas2017,Tremaine2023}.

A truly efficient simulation method for a gravitational system demands adaptation of time-scales of the method.  So-called hybrid methods \citep{C99,K00,Oshino2011,GrimmStadel2014,Iwasawaetal2017,Wisdom2017,Reinetal2019,Lu2024} achieve this in a rather limited and blunt, but effective, way by changing the Hamiltonian we are solving.  A more sophisticated--and, perhaps, complete--method is provided by \cite{DLL98} (see also \citealt{SB94, Leeetal1997}), in which, depending on a pair of particles' separation, a different time-scale is used to solve every (non-Solar) particle pair.  One might argue this approach is a more efficient use of compute resources.  Indeed, \cite{MakinoHut1988} show, analytically and through simulation, that the more fine-grained the time-scale control in $N$-body simulations is, the more efficient the method is.  The \cite{DLL98} algorithm has been parallelized \citep{Wishardetal2020}.

\cite{HD2024}, hereafter HD24, developed the \cite{DLL98} approach further by explicitly assigning timesteps to each particle pair, which themselves can vary in time independently\footnote{This wording is not exactly precise, as one pair's timestep changes can affect other pairs' timesteps to a small degree.}.  If $N$ is the number of planets, the number of timesteps is $N(N-1)/2$.  This discrete variation can be accomplished time-reversibly, abandoning the need for the strict rules of symplectic methods, while preserving most of their benefits \citep{HMM95,Funatoetal96,quinn97,Makinoetal2006,pelu12,D17,HB18}.  The HD24 algorithms had a limitation in that pairwise steps could only change before and after a specified global timestep, but not during the step.  While the algorithm performed well, this limitation certainly reduced efficiency: a pair undergoing an extreme close encounter needed a small timestep for the entire global step, even when the encounter occurred during a fraction of the global step.  For example, for $e = 0.999$, about $4\times 10^{6}$ small steps in a global step might be required, but only a fraction of those small steps are actually needed to resolve the pericentre passage.

In this work, we significantly improve the efficiency of the HD24 algorithm by allowing timestep adaptation on the fly, independent of the global stepsize.  This improvement is inspired by \cite{McMillan1986,Makino1991mod}.  We call our implementation in \texttt{C} ``\texttt{MTR}'' for Multiple Timestep Reversible.  \texttt{MTR} has been translated to \texttt{C} by \texttt{Claude} AI from a code written by the authors in \texttt{MATLAB}.  The theory describing it is in Sections \ref{sec:math} and \ref{sec:codelog}.  We carefully test that the code exhibits no secular drift in energy error, as required of a time-symmetric code, in Section \ref{sec:tests}.  We also explicitly test its time symmetry, using round-trip tests, in this section.

Finally, in Section \ref{sec:Moon}, we compare the performance of \texttt{MTR} against the state-of-the-art \texttt{GPLUM} \citep{Ishigaki2021} code by testing an accretion of the Moon problem.  The number of particles is varied, and \texttt{MTR} is found to be up to two orders of magnitude faster than \texttt{GPLUM}'s Hermite mode (for which it is not optimized), and our default \texttt{MTR} configuration matches or exceeds the \texttt{GPLUM} tree mode, despite \texttt{GPLUM} only approximating long-range forces.  A discussion is provided in Section \ref{sec:conc}.  For reference, a table of the main symbols and functions in this work is provided in Table \ref{tab:symbols}.

%%%%%%%%%%%%%%%%%%%%%%%%%%%%%%%%%%%%%%%%%%%%%%%%%%%%%%%%%%%%
\section{Mathematical background}
\label{sec:math}
We begin by writing the $N$-body Hamiltonian in Democratic Heliocentric Coordinates (DHC) \citep{DLL98}.  They are useful for gravitational dynamics problems with a dominant mass in which motion around it is near-Keplerian.  We call the two types of particles ``star'' and ``planet.''  Written as $(\v{Q},\v{P})$, these coordinates conveniently separate out the centre-of-mass degrees of freedom, which can then be ignored by shifting to a barycentric frame.  Write the Hamiltonian as $H = H_{\mathrm{Kep}} + H_{\mathrm{Sun}}  + V $, where $H_{\mathrm{Kep}}$ is a sum of independent Keplerians, $H_{\mathrm{Sun}} $ is the stellar kinetic energy, and $V$ is the planet--planet potential:
\begin{equation}
\label{eq:Hdef}
\begin{aligned}
H_{\mathrm{Kep}} =& \sum_{i \ne 0} H_{\mathrm{Kep},i} = \sum_{i \ne 0} \left(\frac{P_i^2}{2 m_i} - \frac{G m_i m_0}{Q_i}  \right), \\
H_{\mathrm{Sun}} =& \frac{1}{2 m_0} \left( \sum_{i \ne 0} P_i \right)^2, \text{and} \\
V =& - \sum_{0 < i < j} \frac{G m_i m_j}{Q_{i j}}.
\end{aligned}
\end{equation}
Here, $m_{i \ne 0}$ is the mass of planet $i$, $m_0$ is the stellar mass, and $\v{Q}_{ij} = \v{Q}_i - \v{Q}_j$.  $\v{P}_{i \ne 0}$ is the barycentric momentum and $\v{Q}_{i \ne 0}$ is the heliocentric position.  It is useful to decompose $H$ into $H_{\mathrm{Kep}}$, $H_{\mathrm{Sun}}$, and $V$, because each is an easily solved Hamiltonian.  This is the basis of the Wisdom--Holman map.  For timestep $h_0$, write it as map $\psi_{h_0}$:
\begin{equation}
\psi_{h_0} = \exp \left(\frac{h_0}{2}  \hat{H}_{\mathrm{Sun}} \right) \exp\left(\frac{h_0}{2}  \hat{V}_{} \right) \exp\left(h_0 \hat{H}_{\mathrm{Kep}} \right)  \exp\left(\frac{h_0}{2}  \hat{V}_{} \right)  \exp\left(\frac{h_0}{2} \hat{H}_{\mathrm{Sun}}\right).
\label{eq:psi2}
\end{equation}
In words, from right to left, $\psi_{h_0}$ evolves orbits according to $\sub{H}{Sun}$ for $h_0/2$, then $V$ for $h_0/2$, $\sub{H}{Kep}$ for $h_0$, $V$ again for $h_0/2$, and finally $\sub{H}{Sun}$ for $h_0/2$.  $\hat{ }~$ denotes an operator with definition, say, $\sub{\hat{H}}{Sun} \v{z} = \{ \v{z}, \sub{H}{Sun} \}$, where $\v{z}$ is the phase space and $\{\}$ denote Poisson brackets.  $\psi_{h_0} $ is a good approximation to solving $H$ for $h_0$ small: its accuracy is $\epsilon h_0^2$, with $\epsilon$ the maximum ratio of planetary to stellar mass.  We will be applying a variation of $\psi_{h_0}$ in which each pair in $V$ gets its own timestep, which can vary in a time-symmetric manner; that is, the same sequence of steps is obtained forwards and backwards in time.

To see how this works, we first consider a two-planet system with a single timestep (because there is only one pair of planets), which we allow to vary.  The planetary separation is denoted $\v{Q}$.  We write 
\begin{equation}
\begin{aligned}
\psi(h_0) &= \exp \left(\frac{h_0}{2}  \hat{H}_{\mathrm{Sun}} \right) \times \\
& \exp\left(\frac{s_0}{2}  \hat{V} \right) \exp\left(s_0 \hat{H}_{\mathrm{Kep}} \right)  \exp\left(\frac{s_0}{2}  \hat{V}_{} \right) \times \\
& \exp\left(\frac{s_1}{2}  \hat{V} \right) \exp\left(s_1 \hat{H}_{\mathrm{Kep}} \right)  \exp\left(\frac{s_1}{2}  \hat{V}_{} \right) \times \\
&\vdots \\
& \exp\left(\frac{s_K}{2}  \hat{V} \right) \exp\left(s_K \hat{H}_{\mathrm{Kep}} \right)  \exp\left(\frac{s_K}{2}  \hat{V}_{} \right) \times \\
&  \exp\left(\frac{h_0}{2} \hat{H}_{\mathrm{Sun}}\right),
\end{aligned}
\label{eq:adap}
\end{equation}
where the $s_i < h_0$ are a sequence of substeps.  In order for $\psi({h_0})$ to approximate the evolution of $H$ for $h_0$, we are required to have $\sum_i s_i = h_0$.  What we are doing here is different from the Adaptive Global (\texttt{AG}) time-reversible algorithm from HD24: for \texttt{AG}, $h_0$ is allowed to vary.  In the course of the step, the sequence $s_i$ is chosen according to a criterion.  We define the allowed timesteps by $h_i/h_{i+1} = M$, with $M$ an integer $> 1$.  Typically we use $M=3$ or $M=4$.  

We describe the simplest procedure for choosing a timestep first, based only on separation.  This is for demonstration purposes only.  Our actual timesteps in this paper will be chosen according to the criteria in Section \ref{sec:tcrit}.  To determine the sequence $s_i$, select cutoff radii 
\begin{equation}
r_1 > r_2 > \hdots > r_L, 
\label{eq:rshell}
\end{equation}
with $L$ a maximum shell number set by the user, and $r_i/r_{i+1} = R$ for $i \ge 0$.  $r_0$ is set by the user (note $r_0 \ge Q > r_1$ is not a relevant range).  $R$ is a real number $> 1$.  The combination of $R$ and $M$ is important, and $\ln M/\ln R$ determines how the timestep scales with separation.  If $Q > r_1$, $s = h_0$; if $r_1 \ge Q > r_2$, $s = h_1$, and so forth.  Before applying each
\begin{equation}
\label{eq:phi}
\phi(s_i) = \exp\left(\frac{s_i}{2}  \hat{V} \right) \exp\left(s_i \hat{H}_{\mathrm{Kep}} \right)  \exp\left(\frac{s_i}{2}  \hat{V}_{} \right),
\end{equation}
we determine what $s_i$ should be, based on how close the planets are; the closer they are, the smaller the timestep.  Before applying $\phi(s_0)$, we assign $s_0$.  After applying $\phi(s_0)$, calculate $s_1$.  If $s_1 < s_0$, redo $\phi(s_0)$, but set $s_0 = s_1$.  This process should be repeated until $s_1$ is no longer smaller than $s_0$.  Then, we can proceed with $\phi(s_1)$.  What we have described here is the (almost) exactly time-reversible \texttt{mapReversibleAnd} algorithm described in \cite{HD2023}, Listing 2, in the case of a timestep that is decreased.

Consider the case when $s_{i+1} > s_i$. $\phi(s_{i+1})$ may be allowed to use this larger $s_{i+1}$ or it may need to continue to use $s_{i+1} = s_i$.  In other words, we may not always allow timestep increases.  The reason is that $\phi(s_i)$ solves a Hamiltonian $H_1$, whose form is not relevant for this work, for time $s_i$.  \texttt{mapReversibleAnd} requires comparing two maps integrating to the same time.  In the case of $h_1$ attempting to increase to $h_0$, with no smaller steps before that, it can only be allowed if $\phi(h_1)$ is applied $M$ times.  For $M = 3$, $\left[\phi (h_1)\phi (h_1) \phi (h_1) \right]$ and $\phi(h_0)$ are the two maps we are switching between reversibly.  Thus, in this case, we are not allowed to increase timesteps after applying only   $[\phi (h_1)]^2$.

It follows that more complicated switches, with timestep increases, are allowed.  Let $t_i = \sum_{j = 0}^{j = i} s_j$ and $s_i = h_l$ for some $l$.  If $t_i/h_l = k M$, with  $k$ an integer $\ge 1$, then we are allowed to demote (increase the timestep) to $h_{l - 1}$.  If $t_i/h_l = k M^2$, $h_l$ can be demoted instead to $h_{l - 2}$.  In general, for $j \ge 1$, if $t_i/h_l = k M^j$, $h_l$ can be demoted to $h_{l - j}$.

In the case of more than two planets, each pair of planets gets its own timestep, which can also increase and decrease according to nearly the same rules, which we describe precisely in Section \ref{sec:codelog}.  The rules are not identical because, in rare cases in which a planet is simultaneously involved in more than one close encounter, the evolution of two or more pairs of planets is correlated.  

To understand the multiple-planet logic, begin by defining operators
\begin{equation}
\hat{A}_k =  \exp\left(\frac{h_k}{2} \sum_{(i,j) \in S_k}\hat{{V}}_{ij} \right),  \qquad \mathrm{and} \qquad \hat{B}_k =  \exp\left({h_k  \sum_{i \in S_k^\prime} \hat{H}_{\mathrm{Kep},i}} \right).
\label{eq:opdef}
\end{equation}
$S_k$ is the set of planet pairs with stepsize $h_k$.  $S_k^\prime$ is a $1$-D array of all the indices in $S_k$ minus the indices in $S_j$ for $j > k$.  So, at level $k$, Kepler solvers are applied to particles `kicked' by $V$ at $k$, but only if they are not kicked at smaller timesteps.  A momentum update by $V$ is typically referred to as a kick.  It follows that $[\hat{B}_k, \hat{A}_j] = 0$ for $j > k$.  $[]$ denotes a commutator.  This important relation implies that the precise location of the Kepler solvers in $\hat{B}_k$ has no impact on the mappings at smaller timesteps, and it is an essential ingredient for ensuring time symmetry of the algorithm.  

The final evolution map is written $\hat{D}_0$:  
\begin{equation}
\begin{aligned}
\hat{D}_{i_{\mathrm{max}}} &=  \hat{A}_{i_{\mathrm{max}}} \hat{B}_{i_{\mathrm{max}}} \hat{A}_{i_{\mathrm{max}}}      ,\\
\hat{D}_i &= \hat{A}_{i}  \hat{D}_{i+1}^M \hat{B}_i \hat{A}_{i}      \text{  for $0 < i < {i_{\mathrm{max}}}$, and}\\
\hat{D}_0 &= \exp\left(\frac{h_0}{2} \hat{H}_\mathrm{Sun}\right) \hat{A}_0 \hat{D}_1^M \hat{B}_0 \hat{A}_0 \exp\left(\frac{h_0}{2} \hat{H}_\mathrm{Sun}\right),
\end{aligned}
\label{eq:Cmap}
\end{equation}
where we defined $\hat{D}_i$ for $0 < i < \sub{i}{max}$ and $\hat{D}_{i_{\mathrm{max}}}$.  $\sub{i}{max}$ labels the largest timestep level in use.  $\hat{D}_0$ is our \texttt{MTR} map, and the logic for increasing and decreasing timesteps is described in Section \ref{sec:codelog}.
%%%%%%%%%%%%%%%%%%%%%%%%%%%%%%%%%%%%%%%%%%%%%%
\subsection{Timestep criterion}
\label{sec:tcrit}
Below Equation \eqref{eq:adap}, we described the timestep criterion based on planet separations, which are compared to cutoff radii $r_i$.  We now choose different criteria.  We are free to do so as long as our criteria are invariant under a change of the sign of time to ensure time symmetry.  We use two criteria in this paper.  First, the $r_i$ in Equation \eqref{eq:rshell} now are set as non-dimensional numbers, and instead of comparing $r_i$ to $Q_{jk}$, we can compare against,
\begin{enumerate}
\item
$Q_{ij}/\sub{r}{Hill}$, where we normalize the separation by some length-scale or Hill radius, or
\label{en:crit1}
\item
\begin{equation}
\label{en}
\frac{\sqrt{\frac{Q_{ij}^3}{G (m_i + m_j)}}}{h_0},
\end{equation}  

\end{enumerate}
a free-fall time-scale normalized by the global stepsize.  This is useful when the Hill radius is poorly defined.  In this criterion, the timestep scales with some power of the free-fall time (depending on $R$ and $M$).

%%%%%%%%%%%%%%%%%%%%%%%%%%%%%%%%%%%%%%%%%%

\section{Description of the code and its logic}
\label{sec:codelog}
Based on one of the criteria in Section \ref{sec:tcrit}, timestep levels are assigned before $\hat{D}_0$.  Note that we are safe to do this as $\sub{\hat{H}}{Sun}$ does not affect the relative planetary separations or velocities upon which sensible timestep criteria (even beyond those in Section \ref{sec:tcrit}) would be built.    A matrix \texttt{levc} is built, and $\texttt{levc}(i,j) = l$ indicates that pair $(i,j)$ is at timestep level $l$.    \texttt{levc} is symmetric and $\texttt{levc}(i,i) = 0$.  $l$ takes values from $0$ (corresponding to $h_0$) to a user-set limit such as $29$.  This limit is $L$ in Equation \eqref{eq:rshell}.  

\texttt{arru} determines which pairs get solved at $\hat{D}_i$ and varies when timestep promotions and demotions occur.  It is an $n \times 2$ array of pairs: $\texttt{arru}(k,:) = (i_k,j_k)$.  At $h_0$, \texttt{arru} contains all unique planet pairs and $n = N(N-1)/2$, with $N$ the number of planets.  At higher levels, \texttt{arru} is filtered and narrowed down (this filtered version is never passed to lower levels). \texttt{arru} is filtered to the pairs at levels $\ge l$, where $l$ is the current level.  The actual pairs mapped at level $l$ are $\mathcal{P}_l$, the pairs at level $l$.  The pairs that actually get kicked are
\begin{equation}
\texttt{indv} = \mathcal{P}_l \cap \texttt{arru}.
\end{equation}
The necessity of this form will become clear below.  The level at which particle $i$ gets mapped with the Kepler solver is given by
\begin{equation}
l_i = \max \{\texttt{levc}(i,j) : (i,j) \in \texttt{arru} \},
\end{equation}
meaning that we find the maximum over all $j$ columns, and only pairs in \texttt{arru} are considered.  An adaptation of the Kepler solver in \cite{WH15} is used.

%%%%%%%%%%%%%%%%%%%%%%%%%%%%%%%%%%%%%%%%%%%%%%%
\subsection{Promotion and demotion of pair timestep levels}
If we are at level $i > 0$ in \texttt{MTR}, we apply the operator (Equation \ref{eq:Cmap}),
\begin{equation}
\hat{D}_i =  \hat{A}_{i}  \hat{D}_{i+1}^M \hat{B}_i \hat{A}_{i}, 
\end{equation}
only if $i < \max ( \texttt{levc})$ at the start of the level.  Otherwise, $\hat{D}_{\sub{i}{max}}$ is applied.  For $i=0$, $\hat{D}_0$ is applied and $\hat{D}_1$ is entered only if $0 < \max (\texttt{levc})$. 

At the start and end of $\hat{D}_i$, for any $i$, we have phase space $(\v{Q},\v{v})$ and $(\v{Q}^ \prime,\v{v}^\prime)$, respectively.  They are used to construct \texttt{levc} and \texttt{levcp}, respectively.  From these, we calculate the difference matrix, $\texttt{F}$,
\begin{equation}
\texttt{F}(i,j) = \texttt{levcp}(i,j) - \texttt{levc}(i,j).
\end{equation}
 %%%%%%%%%%%%%%%%%%%%%%%%%%%%%%%%%%%%%%%%%%
\subsubsection{Promotion}
\label{sec:prom}
Promotion is triggered if any $\texttt{F}$ element is positive.  By promotion, we mean the level is increased, timestep is reduced, and at least one pair must redo kick and Kepler solver maps.  The set of pairs to be repeated may be greater than the set for which $\texttt{F} > 0$.  The substep is repeated and \texttt{arru} determines the repeated set of pairs; it is, 
\begin{equation}
\texttt{arru} \leftarrow \{(i,j) \in \texttt{arru} : \texttt{levc}(i,j) \ge l\ \text{and} \ (i,j) \ \text{overlaps promoted pairs}\}.
\label{eq:arrunarr}
\end{equation}
In other words, the repeated set $\mathcal S$ consists of all pairs which have overlap with the promoted pairs and whose levels are greater than or equal to the current level.  ``Overlaps'' needs to be defined carefully.  We use an example to demonstrate.  If $(1,5)$ gets repeated at level $2$, at some substep, all pairs involving planets $1$ or $5$ and whose levels are $\ge 2$ must be repeated and added to $\mathcal S$; call this new set $\mathcal S_1$.  Then, all pairs which share a member with $\mathcal S_1$ must be included in $\mathcal S$ as well: call this third set $\mathcal S_2$.  This chain must be continued until closure and we obtain our final set $\mathcal S$.  In practice, usually $(1,5)$ would be repeated alone unless one of these planets is involved in simultaneous close encounters with two or more other planets.

Set $S$ has its phase space reset from $(\v{Q}^\prime,\v{v}^\prime)$ to $(\v{Q},\v{v})$.  \texttt{levc} is replaced,
\begin{equation}
\texttt{levc}(i,j) \leftarrow \texttt{levcp}(i,j),
\end{equation}
and repetition can then proceed.  Note that repetition can occur more than once if $\texttt{levcp} - \texttt{levc}$ still has array elements $ > 0$.
%%%%%%%%%%%%%%%%%%%%%%%%%%%%%%%%%%%%%%%%%%%%%%
\subsubsection{Demotion}
\label{sec:demo}
Demotions are not always allowed, as described in Section \ref{sec:math}.  This is to ensure reversibility.  A pair is eligible for demotion if $\texttt{F}(i,j) < 0$.  Demotions are only allowed at the last substep of $\hat{D}_i$, and only the value of $\texttt{F}(i,j)$ at this last substep matters.

%%%%%%%%%%%%%%%%%%%%%%%%%%%%%%%%%%%%%%%%%%%%%%%%%%%
%%%%%%%%%%%%%%%%%%%%%%%%%%%%%%%%%%%%%%%%%%%%%%
\subsubsection{Mixed promotion/demotion}
\label{sec:mixed}
If a system of particles is large, say in a planetesimal system with $10^ 4$ particles, it is likely that a substep can have a set $\mathcal S$ getting promoted while a set $\mathcal T$ of pairs is demoted.  In most cases, there is no particle overlap between $\mathcal S$ and $\mathcal T$ and both promotion and demotion proceed normally and simultaneously.

A new case, again requiring that a planet is involved in multiple close encounters simultaneously, is when at least one planet is in both $\mathcal S$ and $\mathcal T$.  Call the set of pairs in $\mathcal T$ with such overlap $\mathcal T^\prime$.  Pairs in $\mathcal T^\prime$ cannot be demoted and instead their substep at level $i$ will be repeated.  This is because the evolutions of the $\mathcal T^\prime$ and $S$ planets are not independent, and after repetition $\mathcal T^\prime$ may no longer be slated for demotion.  The code is prepared for mixed promotion and demotion with overlap, and, indeed, such cases are a common occurrence in our tests.  For example, in all of the $975$-body simulations of Section \ref{sec:Moon}, every one of the $95$ global timesteps had a substep where a particle was simultaneously slated for promotion and demotion.  

During repetition, \texttt{arru} is narrowed, as described by Equation \eqref{eq:arrunarr}.  When repetitions are complete, \texttt{arru} must again be broadened.  \texttt{levc} has now been updated, and we simply set \texttt{arru} to pairs for which $\texttt{levc} \ge l$, with $l$ the current level.

%%%%%%%%%%%%%%%%%%%%%%%%%%%%%%%%%%%%%%%%%%%%%%%%%%%%
\subsection{Example global step}
\label{sec:exam}
\begin{figure}
    \centering
    \includegraphics[width=0.5\textwidth]{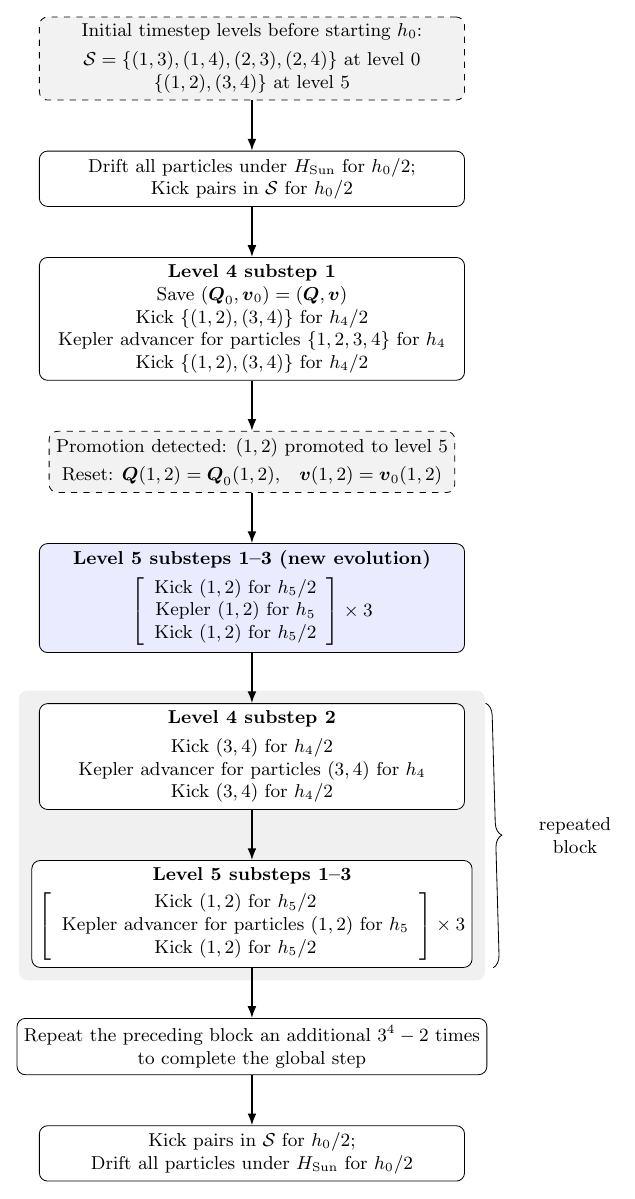}
    \caption{An example fictitious global timestep $h_0$ for a four-planet double-binary system in which a pairwise planetary timestep requires repetition.  There are $6$ pairs of planets and thus $6$ timesteps (all of which can vary in time), with the two binary pairs denoted as $(1,2)$ and $(3,4)$.  All pairs except these two are at large timesteps.  The two binaries initially have small timestep $h_4 = h_0/3^4$, and thus require $3^4$ substeps per global step.  Because the binaries are eccentric, promotion to $h_5$ is needed for pair $(1,2)$ in this global timestep.  Thus, its first substep is repeated to ensure time symmetry.  Only one substep out of $3^5 + 1$ for $(1,2)$ is repeated.
    }
    \label{fig:diagram}
\end{figure}

We show how promotion might work in practice during a step for a five-body problem consisting of two binaries orbiting a star.  This is similar to the test problem of Section \ref{sec:doubbin}, which has frequent timestep promotions and demotions for the two binary pairs.  An example timestep workflow is shown in Fig. \ref{fig:diagram}.  
%%%%%%%%%%%%%%%%%%%%%%%%%%%%%%%%%%%%%%%%%%%%%%%%%%%
%%%%%%%%%%%%%%%%%%%%%%%%%%%%%%%%%%%%%%%%%%%%%%%%%%%%%
For four planets, there are six pairs of planets, and thus six timesteps varying in time.  The two binary pairs have the smallest timesteps, and the other four pairs remain at level $0$: they are called set $S$.  Timesteps are selected based on the free-fall criterion.  The workflow proceeds from top to bottom and is described as follows.
\begin{enumerate}
\item
Positions of all particles are updated according to $\sub{H}{Sun}$.
\item
Pairs in $S$ are kicked at level $0$.
\item
We enter level $4$ and save the state $(\v{Q}_0,\v{v}_0)$ (the state is saved before any substep, including the level-$0$ substep).  $M = 3$, so three substeps are taken per level.  $\sub{\hat{H}}{Sun}$ is always applied exactly twice.  A substep is defined as a sequence of kick--Kepler--kick applied to a pair of particles.  The Kepler level may be higher than the kick level, and this still defines a substep.  Here, a substep is, for instance, a kick to $(1,2)$, Kepler solver to $1$ and $2$, and kick again to $(1,2)$, all at level $5$. 
\item
In the first substep, $(1,2)$ and $(3,4)$ are kicked, a Kepler advancer is applied, and then a new kick is applied.
\item
Promotion is detected for $(1,2)$, so their $(\v{Q},\v{v})$ vectors are reset.
\item
The $(1,2)$ substep is now redone at level $5$.  Three substeps at level $5$ are performed.  Repetition is finished.
\item
Level $4$ substep $2$ is carried out, followed by three level-$5$ substeps.
\label{en:sequence}
\item
The sequence \ref{en:sequence} is carried out $3^4 - 2$ times, so that the two pairs $(1,2)$ and $(3,4)$ have been integrated (for a different Hamiltonian from $H$) for $3^4 h_4 = h_0$.
\item
Kick pairs in $S$ and drift all particles under $\sub{H}{Sun}$.  
\end{enumerate}
Note that to make this sequence time-reversible, only one substep out of $3^5+1$ for $(1,2)$ was repeated.
%%%%%%%%%%%%%%%%%%%%%%%%%%%%%%%%%%%%%%%%%%%%%%%
\section{Testing error and speed performance}
\label{sec:tests}
The code in Section \ref{sec:codelog} was originally written in \texttt{MATLAB} by the authors.  \texttt{ChatGPT} aided in efficiency improvements. A translation to \texttt{C} was carried out by \texttt{Claude} AI and guided by the authors.  All aspects of the code were carefully checked by the authors.  This \texttt{C} version is what we denote as \texttt{MTR} henceforth and what we use for testing.  It is available on GitHub\footnote{https://github.com/dmhernan/Multiple-Timestep-Reversible}.

%%%%%%%%%%%%%%%%%%%%%%%%%%%%%%%%%%%%%%%%%%%%%%%%%%%%%%%%%%
\subsection{Double-binary planet}
\label{sec:doubbin}
We study a double-binary planet problem (upon which the example in Section \ref{sec:exam} and Fig. \ref{fig:diagram} are based), which was also solved in HD24.  Two eccentric binary pairs of planets orbit a star; one binary is placed at $1$ au and another at $3$ au.  The motion is regular (evidence for this will be shown in Figs. \ref{fig:qnorm} and \ref{fig:vnorm}).  Due to the binary eccentricities, both change their pairwise timesteps during their periods ($11.4$ and $12.1$ days).  Initially, the inner binary varies by two timestep levels per period (it is more eccentric), while the outer binary varies by one.  

We highlight that codes such as \texttt{MTR}, \texttt{MERCURIUS}, \texttt{TRACE}, and \texttt{SyMBA} are not actually designed for such problems, because they assume primarily Keplerian motion around the star.  However, such codes can still handle this problem, albeit inefficiently, and it is a good test that forces frequent pairwise timestep changes, our main interest here.  Codes and methods exist for systems with hierarchical orbits; see \cite{C02,GBP14,HB15,HP16,Agoletal2021}.

Both binaries are placed at apocentre initially, but go out of phase as their mutual semi-major axes are different.  This allows us to test the mixed promotion and demotion timestep logic (Section \ref{sec:mixed}).  The inner binary has $a = 0.0125$ au and eccentricity $e = 0.6$, while the outer binary has $a = 0.013$ au and $e = 0.2$.  The masses of all planets are $10^{-3} M_{\odot}$, while the stellar mass is $1 M_{\odot}$.  We set $h_0 = 0.01$ yr and integrate for $1000$ yr so that the inner and outer binaries have 32,000 and 30,171 mutual orbits.  The inner and outer binaries orbit the star $1000$ and $192$ times, respectively.  We set $R = 2$, $M = 3$, $r_0 = 30$, and use the free-fall timestep criterion \ref{en}.

\begin{figure}
    \centering
    \includegraphics[width=0.5\textwidth]{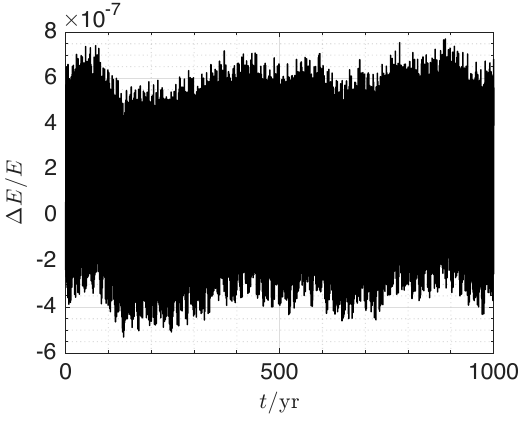}
    \caption{ Energy error versus time for a double-binary planet problem, from Section \ref{sec:doubbin}, solved with \texttt{MTR}.  The two binaries have periods of $11.4$ and $12.1$ days, and their timestep levels vary by up to three values per period.  We integrate for $1000$ yr, making this a challenging test with frequent timestep changes.  There is no secular increase in the energy error.  If the timestep changes were irreversible, the error would rapidly accumulate. 
    }
    \label{fig:dEbin}
\end{figure}

\begin{figure}
    \centering
    \includegraphics[width=0.5\textwidth]{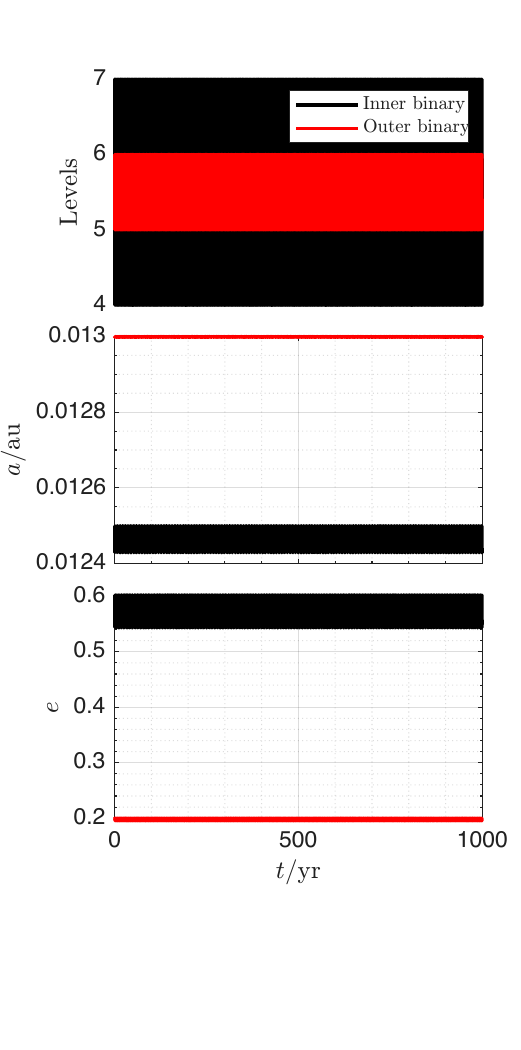}
    \caption{Evolution in time of the timesteps, semi-major axes, and eccentricities of the two binary pairs in the double-binary problem of Section \ref{sec:doubbin}.  The frequent timestep changes (measured between global timesteps) make the top subpanel completely shaded.  One binary's timesteps vary between four values, while the other's vary between two values.  The variations in eccentricities and semi-major axes are bounded, as expected of a non-chaotic problem.
        }
    \label{fig:binorb}
\end{figure}

Fig. \ref{fig:dEbin} shows the energy error over time.  The evolution over time of the binary timestep levels, their mutual semi-major axes, and their eccentricities is shown in Fig. \ref{fig:binorb}.  Data is plotted at every global step.  
The timestep levels are changed so frequently that its figure is fully shaded.  Initially, the inner binary oscillates only between levels $4$ and $6$ during a period, but it then oscillates between levels $5$ and $7$ as the orbits change.  

Despite the frequent timestep switches, the energy shows no secular increase in time, with the absolute error remaining below $10^{-6}$.  By calculating the average error every $100$ steps (not shown), we verified that there is no secular error drift masked by the large error amplitudes in Fig. \ref{fig:dEbin}.  We can contrast this with the results of HD24, who test a naive, irreversible, adaptive timestepping method on an almost identical problem in the red curve of their Fig. 6.  Over only a $100$ yr integration time, their naive method's absolute error grows by about four orders of magnitude.  {We also verified \texttt{MTR}'s absolute energy error scales as $h_0^2$ for this problem.  This is predicted by the analysis in the text under Equation (36) in HD24, which approximately applies here.}

The total number of substeps taken was 94,553,226, with the number of repeated steps 126,394, for a repetition fraction of $0.001337$.  $60$ substeps with mixed promotion and demotion were performed (note such substeps are still reversible).  The repeated-substep fractions for inner and outer binary, respectively, were $0.00177$ and $0.00076$.  All other pairs remained at $h_0$.  By construction, the code conserves angular momentum to roundoff error.  In this case the relative error is $-4.4 \times 10^{-13}$, which is an advantage over other general-purpose $N$-body codes.  The orbital elements remain stable over time; they have bounded variation for both the inner and outer binary.  

\begin{figure}
    \centering
    \includegraphics[width=0.5\textwidth]{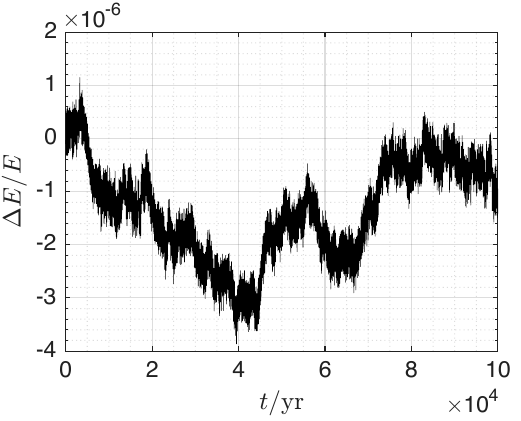}
    \caption{ Same as Fig. \ref{fig:dEbin}, but the integration time has been extended by a factor of $100$, while energy error is only plotted every $10^3$ global steps.  There is no secular energy trend.
        }
    \label{fig:binorblong}
\end{figure}

We have also verified a lack of secular energy changes over longer time-scales.  Fig. \ref{fig:binorblong} is the same as Fig. \ref{fig:dEbin}, but the integration time has been extended to $10^5$ yr, and output is now only plotted every $10^3$ global steps.
%%%%%%%%%%%%%%%%%%%%%%%%%%%%%%%%%%%%%%%%%%%%%%%%%%%%%
\subsection{Verifying time symmetry}
A test exists to verify that our double-binary integration is actually time-reversible.  Begin with initial conditions $(\v{Q}_0,\v{v}_0)$.  All parameters are the same as in Section \ref{sec:doubbin}.  Integrate for a number of global timesteps $m$.  \texttt{levc} is saved.  Flip the sign of velocities and again integrate for $m$ steps, using \texttt{levc} to determine the starting timesteps.  Note that the timestep criterion is required to be invariant under a change of the sign of time, so that \texttt{levc} is unchanged under the velocity flips.  After the forwards and backwards integrations, flip the velocities once more, and save the state as $(\v{Q}_{f},\v{v}_{f})$.  An $L_2$ norm is then computed between the two state vectors:
\begin{equation}
\label{eq:norm}
\begin{aligned}
\sub{Q}{norm} &= \sqrt {\sum_{i > 0} \left(\v{Q}_{{i},f} - \v{Q}_{i,0} \right)^2}, \qquad \text{and} \\
\sub{v}{norm} &= \sqrt{\sum_i \left(\v{v}_{{i},f} - \v{v}_{i,0} \right)^2 }, \\
\end{aligned}
\end{equation}
where $\v{Q}_{i,0}$, $\v{Q}_{{i},f}$, $\v{v}_{i,0}$, and $\v{v}_{{i},f}$ refer to particle $i$.  Note that the sum in $\sub{Q}{norm}$ is over $i > 0$, since the $0$ index is the irrelevant centre of mass, but the sum in $\sub{v}{norm}$ is over all $i$, since we have the stellar state.  We also compute the individual planetary $L_2$ norms with the same restrictions on $i$:
\begin{equation}
\label{eq:normpl}
\begin{aligned}
\v{Q}_{i,\mathrm{norm}} &=  \left |\v{Q}_{{i},f} - \v{Q}_{i,0}  \right|, \qquad \text{and} \\
\v{v}_{i,\mathrm{norm}} &= \left | \v{v}_{{i},f} - \v{v}_{i,0} \right |. \\
\end{aligned}
\end{equation}
This experiment is repeated for $m$ ranging from $1$ to $10^{6}$, or for integration times from $0.01$ yr to $10^4$ yr.  

\begin{figure}
    \centering
    \includegraphics[width=0.5\textwidth]{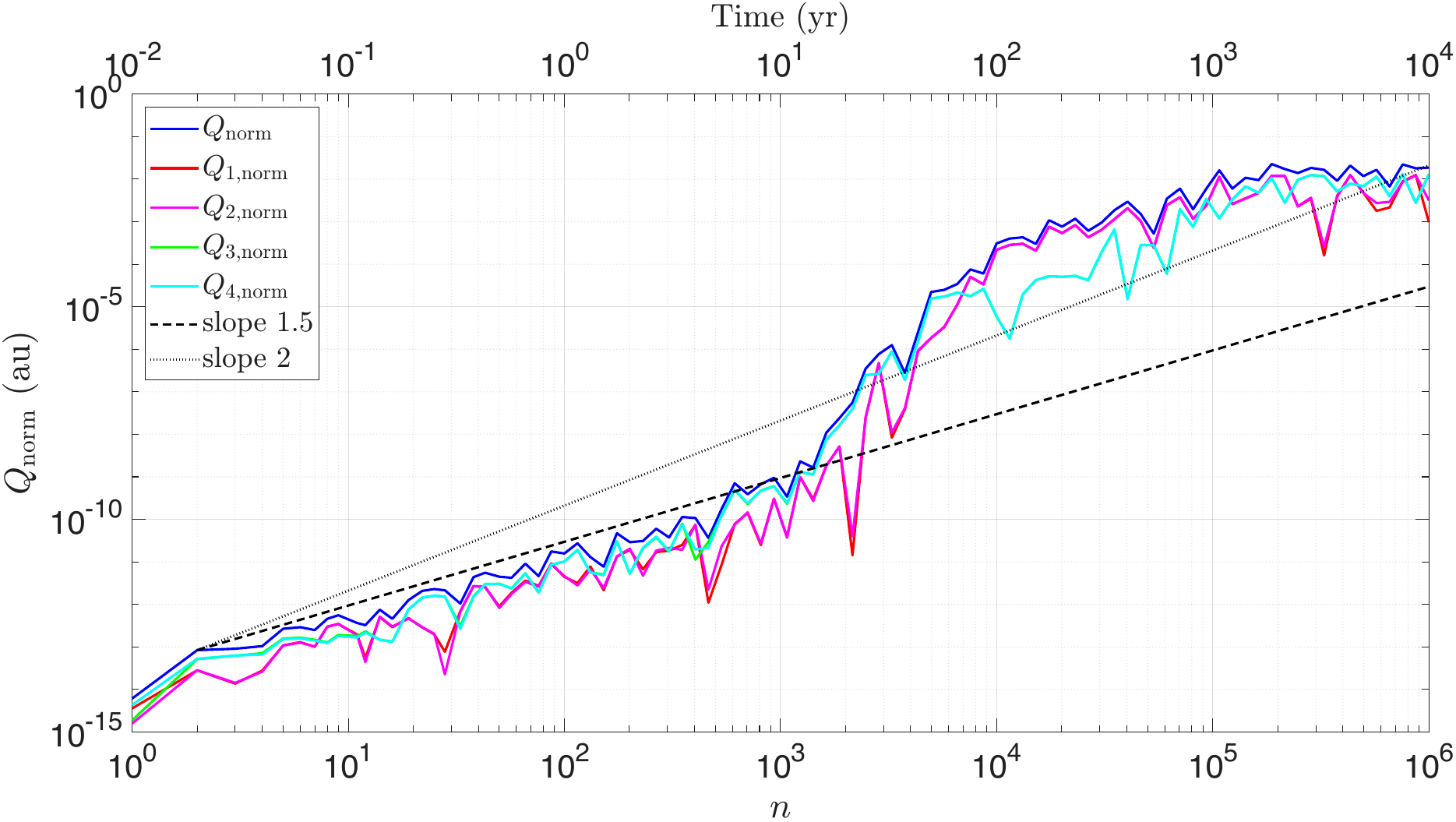}
    \caption{
    Time-symmetry test of \texttt{MTR}.  Using the double-binary problem of Section \ref{sec:doubbin} again, with the same initial conditions and parameters, we integrate for $m$ steps.  We reverse the sign of the velocities, again integrate for $m$ steps, and finally reverse the sign of the velocities again.  We then calculate the $L_2$ norm between the final and initial $(\v{Q},\v{v})$ states.  This experiment is repeated for different $m$.  This plot shows the $L_2$ norm for each planet's $\v{Q}$ as well as the overall $\v{Q}$ norm (Equations \ref{eq:norm} and \ref{eq:normpl}).  For $m < 10^3$ ($< 10$ yr), the error is near machine precision and growing as $m^{3/2}$, which is optimal according to Brouwer's law.  Any irreversible timestep changes would be obvious on this plot as, to take an example, $1265$ substep repetitions occur for the $m = 10^3$ forward integration.  For $m > 10^3$, we have errors due to the non-chaotic divergence of trajectories.  
    }
    \label{fig:qnorm}
\end{figure}

\begin{figure}
    \centering
    \includegraphics[width=0.5\textwidth]{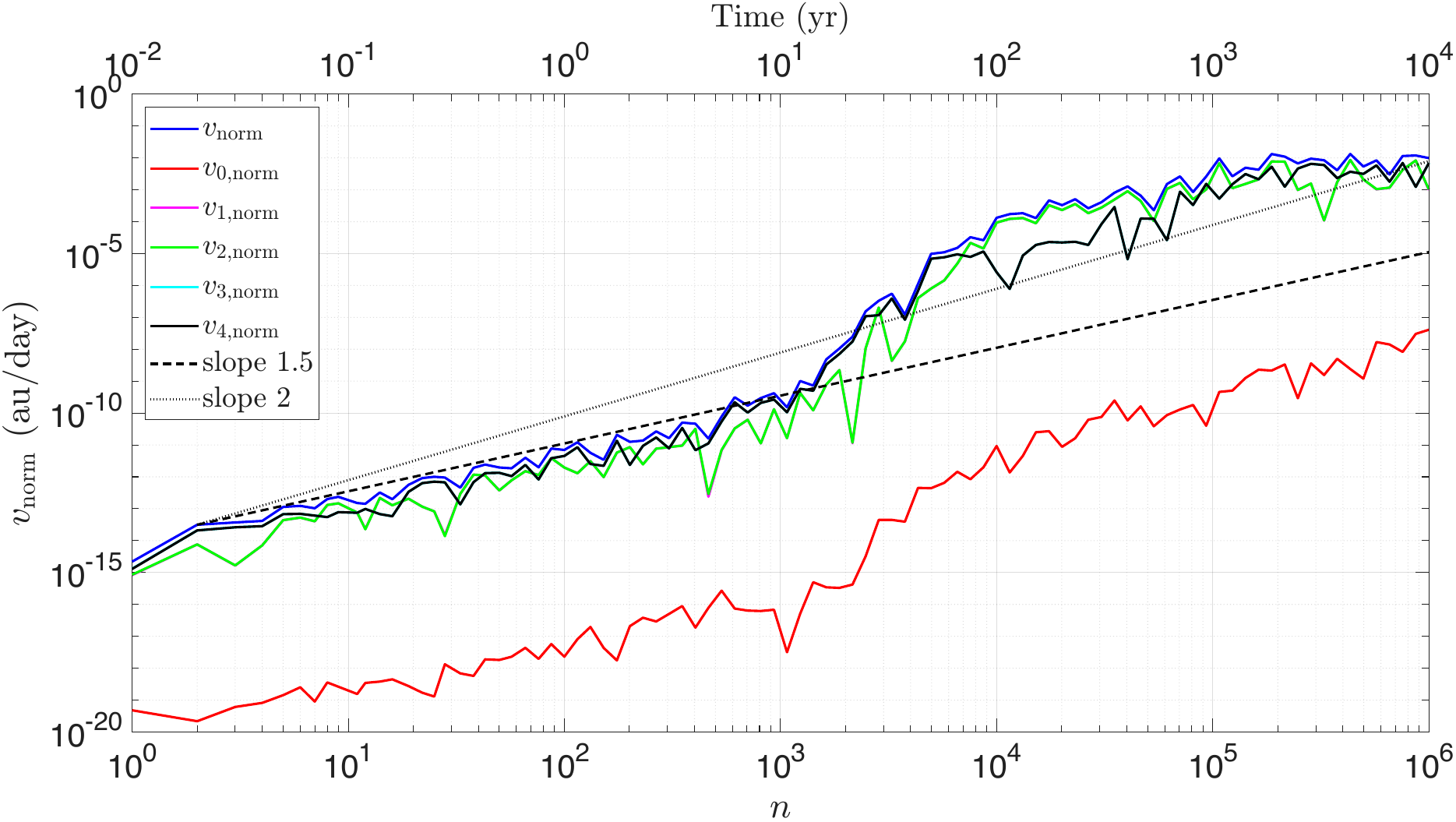}
    \caption{
    Same as Fig. \ref{fig:qnorm}, but now the velocity norms are shown.  We also show the velocity norm of the star.  The behaviours are similar to those in Fig. \ref{fig:qnorm}.
        }
    \label{fig:vnorm}
\end{figure}

This experiment is powerful because \texttt{MTR}'s truncation error is orders of magnitude larger than machine precision, but if \texttt{MTR} is time-reversible, this contribution is wiped out, allowing us to check the accumulation of roundoff error of the method.  The growth of $\sub{Q}{norm}$ in this experiment can also indicate whether the motion is regular or chaotic.  Finally, this round-trip test can quantify the contribution to roundoff error or divergence of trajectories due to each specific planet.  The results for $\sub{Q}{norm}$ and $\sub{v}{norm}$ are shown in Figs. \ref{fig:qnorm} and \ref{fig:vnorm}, respectively.

The plots provide evidence that \texttt{MTR} is time-reversible.  For $m < 10^3$, the error in all norms is near machine precision and growing at a rate of $m^{3/2}$.  According to Brouwer's law, optimal roundoff error growth in angle-like coordinates (like $\v{Q}$ and $\v{v}$) is $(2 m)^{3/2}$, in agreement with our results.  Note that after $101$ steps there are $129$ repeated substeps, and after $10^3$ steps, there are $1265$ repetitions.  There is no evidence of any irreversibilities arising from these pairwise timestep changes.  A slope of $2$, indicating biased growth, is also shown, and no curves follow it.

The curves for $Q_{1,\mathrm{norm}}$ and $Q_{2,\mathrm{norm}}$ largely overlap, as do the $Q_{3,\mathrm{norm}}$ and $Q_{4,\mathrm{norm}}$ curves.  As $(1,2)$ and $(3,4)$ are the binary pairs, it stands to reason that both sets of curves are correlated.  The $\sub{v}{norm}$ curves show the same correlation.  

For $m > 10^3$, the error growth is no longer described by roundoff error.  This error growth rate is due to the differential equations and the physical system itself.  The error growth being slower than exponential is consistent with the system being non-chaotic, as expected from the stable orbital behaviour in Fig. \ref{fig:binorb}.  The error saturates at $\approx 10^{-2}$ au for $\sub{Q}{norm}$ and $\approx 10^{-2}$ au$/$day for $\sub{v}{norm}$; these values agree to within an order of magnitude of the maximum relative speeds and separations of the binaries, which are $1.4 \times 10^{-2}$ au/day and $0.02$ au. Thus, the $\sub{Q}{norm}$ and $\sub{v}{norm}$ saturations in Figs. \ref{fig:qnorm} and \ref{fig:vnorm} are consistent with arising from two states of a regular, non-chaotic system that differ in the phases of the binary orbits.

%%%%%%%%%%%%%%%%%%%%%%%%%%%%%%%%%%%%%%%%%%%%%%%
%%%%%%%%%%%%%%%%%%%%%%%%%%%%%%%%%%%%%%%%%%%%%%%%%%%%%
\subsection{Violent Solar system}
\label{sec:viol}
The violent Solar system problem was introduced by \cite{DLL98} and consists of the Sun and outer giant planets; the planetary masses are scaled up by a factor of $50$ to induce close encounters and planetary scattering.  We use $h_0 = 0.03$ yr and a runtime of $3000$ yr.  We use $r_0 = 3$, and this value decreases by a factor $R = 2$.  We use timestep criterion \ref{en:crit1} from Section \ref{sec:tcrit}, with Hill radius equal to Saturn's, $1.61$ au\footnote{HD24 have a typo in their paper and code, and instead set $\sub{r}{Hill} = 1.52$ au.  This $\sub{r}{Hill}$ made their timesteps larger, but without impact on reversibility or code performance.  Moreover, under their Equation (19), $R = 2$ is correct to one significant figure, but more precisely, $R = 3^{2/3}$.  We use $R = 2$ here.}.  We use $M = 4$.  According to the analysis in HD24, Equation (36), (see also \citealt{DLL98}), this choice of $R$ and $M$ makes the energy error approximately independent of how close the planets approach each other.

Only $40$ substeps are repeated in this problem, making up a fraction $6.5 \times 10^{-5}$.  The difference between a naive explicit method and our reversible method is merely these $40$ substeps.  A close encounter occurs at $283$ yr between Jupiter and Saturn, with the minimum timestep, recorded between global steps (the minimum within global timesteps is possibly less), equal to $h_5 = 2.9 \times 10^{-5}$ yr.  The minimum separation is $0.077$ au.  A close encounter between Uranus and Neptune occurs at $1054$ yr, with minimum timestep $h_3 = 4.7 \times 10^{-4}$ yr and minimum separation $0.36$ au.  Saturn is ejected after $420$ yr. 

\begin{figure}
    \centering
    \includegraphics[width=0.5\textwidth]{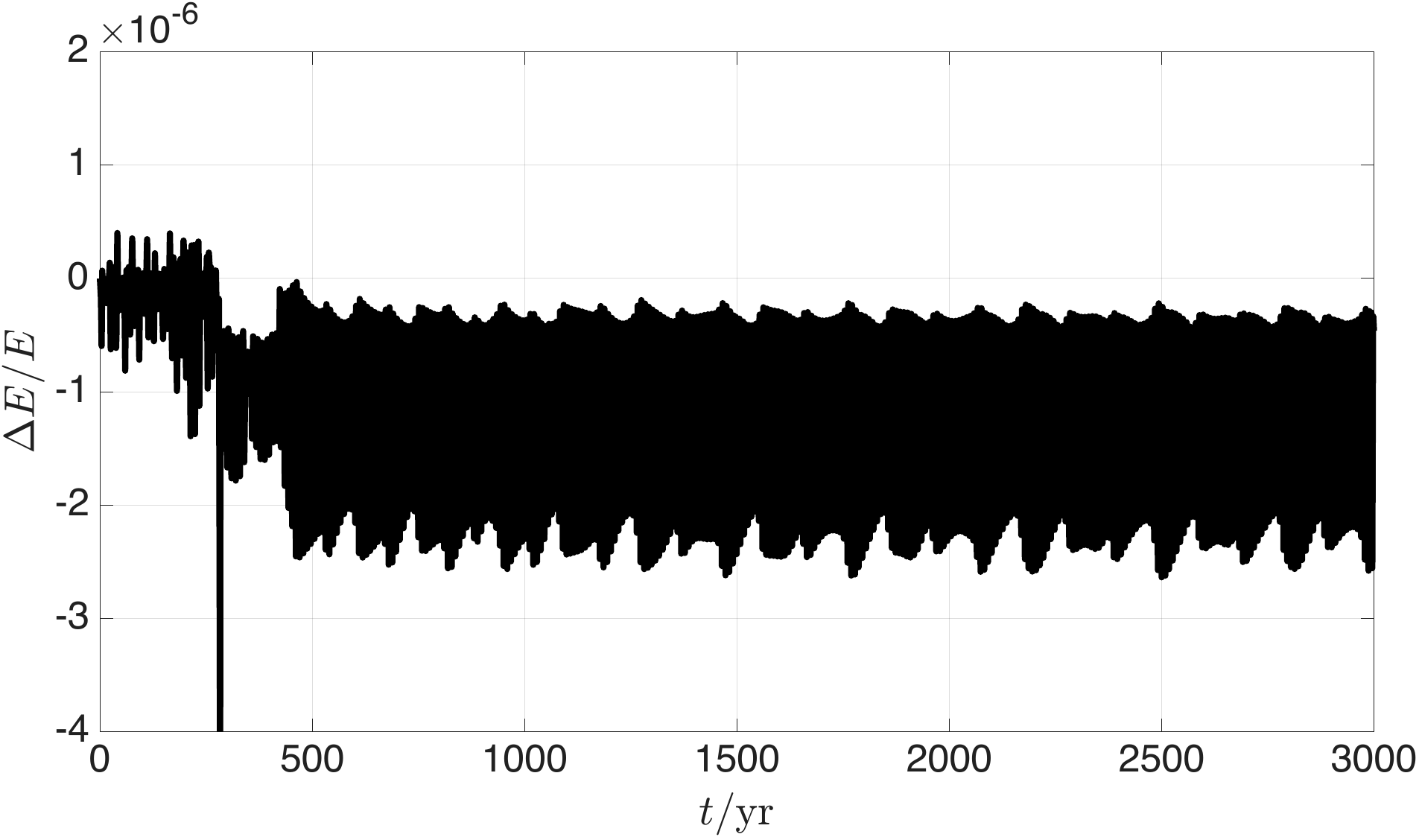}
    \caption{
    Energy error as a function of time for the violent outer Solar system problem of Section \ref{sec:viol}, run with \texttt{MTR}.  A close encounter between Jupiter and Saturn occurs at $283$ yr, while a close Uranus--Neptune encounter occurs at $1054$ yr.  The energy error is small and stable.  Only a fraction $6.5 \times 10^{-5}$ of substeps is repeated, so \texttt{MTR} is behaving essentially at the speed of an explicit method.
    }
    \label{fig:violente}
\end{figure}

We show the energy error over time for this problem in Fig. \ref{fig:violente}.
The behaviour is excellent despite a small error shift occurring at the Jupiter--Saturn close encounter.

%%%%%%%%%%%%%%%%%%%%%%%%%%%%%%%%%%%%%%%%%%%%%%%%%%%%%%%%%%%%%%%%%%%%%%%%%%%%%%%%%%%%%%%
\subsection{The accretion of the Moon}
\label{sec:Moon}
All tests so far involved at most a handful of bodies and isolated close encounters. 
We now want to see how \texttt{MTR} behaves for large gravitational systems, where close encounters are frequent and repeated. 
For this purpose we turn to the accretion of the Moon from an impact-generated debris disc of $\sim10^3$ particles around the Earth, which \citet{Ida1997} simulated with a fourth-order Hermite scheme \citep{Makino1991} with individual timesteps \citep{Aarseth1963,McMillan1986}. 
\citet[Section 6.5]{DLL98} initialized such a disc, turned physical collisions off, and used this system to test the error properties of \texttt{SyMBA}.  \citet{Lu2024} did the same for \texttt{TRACE}. We reproduce this test here, using \texttt{MTR} and the \texttt{GPLUM} code \citep{Ishigaki2021}.  In what follows, we describe the setup of this problem and the results.

%%%%%%%%%%%%%%%%%%%%%%%%%%%%%%%%%%%%%%%%%%%%%%%%%%%%%%%%%%%%%%%%%%%%%%%%%%%%%%%%%%%%%%%%%%%
\subsubsection{Initial conditions} 
\label{sec:Moon:IC}
We follow the setup of \citet[Section 6.5]{DLL98}. 
We use units in which $M_{\oplus}$ = 1, the Roche radius $R_{\rm Roche} = 1$ ($= 2.9R_{\oplus}$), and $G=1$, so that a body with $a=R_{\rm Roche}$ has an orbital period of $2\pi$\footnote{\texttt{MTR} uses $G=2.959\times10^{-4}\,{\rm au^3\,M_\odot^{-1}\,day^{-2}}$. 
We therefore rescale the masses by $1/G$ and leave lengths and times unchanged.}. 
The disc particles have masses between $3.2\times10^{-7} M_{\oplus}$ and $3.2\times10^{-4}\,M_\oplus$, semi-major axes between $0.7\,R_{\rm Roche}$ and $1.3\,R_{\rm Roche}$, and eccentricities and inclinations drawn from Rayleigh distributions with mean 0.3, truncated
at $e=0.95$ and $i=50^\circ$ \citep{Ida1997}.
As in \citet{DLL98}, particles whose initial pericentre lies inside the Earth are removed, which leaves
$N=975$ (non-Solar) particles with a total mass of $0.041\,M_\oplus$, 3.3 lunar masses, orbiting the Earth.  One of our goals is to measure how \texttt{MTR}'s compute effort scales with $N$, so we also integrate the subsets consisting of the first $N=50$, 100, 200, 400, and 800 particles of this disc. Every run is integrated for three orbital periods, $t_{\rm end}=6\pi$, with the global timestep $h_0=0.2$, which corresponds to about 31 steps per orbit at the Roche radius.  The particles are assumed to be point masses without physical collisions.

For \texttt{MTR} we use timestep criterion \ref{en:crit1} of Section \ref{sec:tcrit} with $r_0=3$, $R=2$, and $L=29$ (Equation \ref{eq:rshell}). The scale $r_{\rm Hill}$ in the criterion is set to $f\,r_{\rm H}$, where $f$ is a free factor and $r_{\rm H}$ is the mutual Hill radius, at $a=R_{\rm Roche}$, of the two most massive particles of the disc,
\begin{equation}
  r_{\rm H}=R_{\rm Roche}\left(\frac{m_1+m_2}{3M_\oplus}\right)^{1/3}
  \approx0.06\,R_{\rm Roche} = 0.174 R_{\oplus},
  \label{eq:Moon_rhill}
\end{equation}
with $m_1=m_2=3.2\times10^{-4}\,M_\oplus$. We test $f=1$, 2, 4, 10, and 15 at every $N$ with $M=3$. 
All runs were repeated with $M=4$, for a total of 60 \texttt{MTR} runs. 
Since in most cases $M=4$ only increases the computation time without improving the accuracy (Fig.~\ref{fig:Moon_efficiency}), we mainly present the $M=3$ results below.

To assess the accuracy and speed of \texttt{MTR}, we compare against the state-of-the-art \texttt{GPLUM} code \citep{Ishigaki2021}, an $N$-body code for planetary system formation built on the particle--particle particle--tree (P$^3$T) scheme of \citet{Oshino2011}. In P$^3$T, the forces from the central body and from neighbours within a cutoff radius, of the order of a few Hill radii, are integrated with the fourth-order Hermite scheme \citep{Makino1991} with individual timesteps \citep{Aarseth1963,McMillan1986}, while the forces from more distant particles are evaluated with a Barnes--Hut tree scheme
\citep{Barnes_Hut1986}. Since the force from the central body is in the Hermite part, \texttt{GPLUM} is well suited to a system dominated by the gravity of the central body. The related P$^3$T code \texttt{PeTar} \citep{Wang2020petar} instead targets star clusters, which have no dominant mass.   We run \texttt{GPLUM} in two modes: (a) the default P$^3$T mode, with tree opening angle $\theta=0.5$, and (b) a pure fourth-order Hermite mode, obtained by setting $\theta=0$ and a cutoff radius much larger than the system.
The latter is essentially the method of \citet{Ida1997}. 
In both modes the Hermite timestep parameter is $\eta=0.02$.
In the P$^3$T mode, \texttt{GPLUM} sets the cutoff radius of the $i$th particle to $\max(3\,r_{{\rm H},i},\ 8\,v_{\rm disp}\Delta t_{\rm tree})$, where $r_{{\rm H},i}$ is the Hill radius of the $i$th particle, $v_{\rm disp}$ the random velocity of the disc, and $\Delta t_{\rm tree}$ the tree timestep.  For this hot disc ($e,i\sim0.3$) we use $\Delta t_{\rm tree}=2^{-7}$, which is small enough that the cutoff is set by the Hill radius, as intended.  In the Hermite mode $\Delta t_{\rm tree}$ plays no role.
We note that all \texttt{MTR} and \texttt{GPLUM} runs were performed using a single CPU core of an Apple M1 Pro processor.

\begin{figure}
  \centering
  \includegraphics[width=\columnwidth]{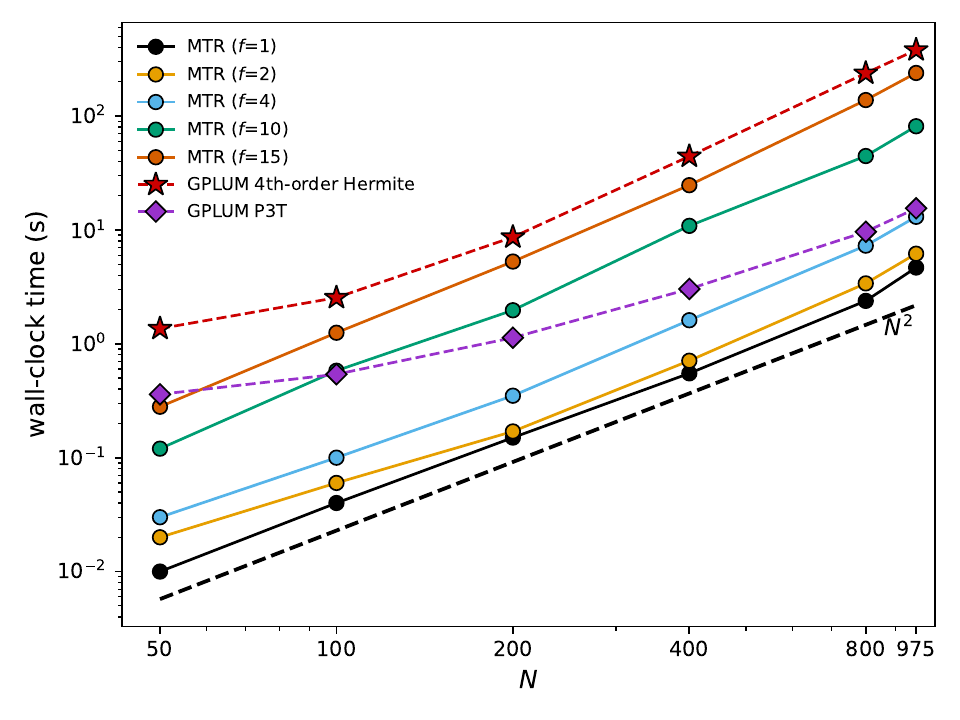}
  \caption{Wall-clock time for three orbital periods of the accretion of the Moon problem as a function of the number of disc particles $N$. Circles show \texttt{MTR} with $M=3$ and $r_{\rm Hill}=f\,r_{\rm H}$ for $f=1$, 2, 4, 10, and 15. Stars and diamonds show \texttt{GPLUM} in the pure fourth-order Hermite and P$^3$T modes, respectively. The dashed line indicates $N^2$ scaling.}
  \label{fig:Moon_walltime}
\end{figure}
%%%%%%%%%%%%%%%%%%%%%%%%%%%%%%%%%%%%%%%%%%%%%%%%%%%%%%%%%%%%%%%%%%%%%%%%%%%%%%%%

\subsubsection{Wall-clock time} 
\label{sec:Moon:wall}
Fig.~\ref{fig:Moon_walltime} shows the wall-clock time of all $M=3$ runs.
The cost of \texttt{MTR} grows with $f$, because more pairs are placed at deep timestep levels.  Relative to $f=1$, the runs with $f=2$, 4, 10, and 15 take about 1.1--2, 2.3--3, 12--20, and 28--58 times longer, respectively.  \texttt{MTR} is faster than the Hermite mode at every $N$ for every $f$.  For $N\ge100$, the speed-up with respect to the Hermite mode is 58--100 for $f=1$, 42--70 for $f=2$, 25--33 for $f=4$, 4--5 for $f=10$, and 1.6--2 for $f=15$.  We note that despite these impressive speedups, \texttt{GPLUM} is not optimized to be run in Hermite mode; it is designed specifically for P$^3$T mode.

\texttt{MTR} with $f\le4$ is also faster than P$^3$T at every $N$.  Both the \texttt{GPLUM} Hermite mode and \texttt{MTR} scale as roughly $N^2$, as expected for direct summation.  \cite{MakinoHut1988} predict scalings as $N^{8/3}$ and $N^{7/3}$ for direct $N$-body codes with shared and individual timesteps, respectively (for a homogeneous density and fixed mass), and \texttt{MTR}'s scalings are somewhat shallower than $N^{7/3}$.  P$^3$T scales as roughly $N^{1.5}$, because direct summation is used only within the Hermite clusters, which in this problem consist of two particles on average, while all other interactions are evaluated by the tree in $\mathcal{O}(N\log N)$.

For large $N$ problems like this one, the number of repetitions may become too large, and we may need to force a stop to these iterations. If a global step still requires repetition after 10 attempts, the code accepts the step as it is\footnote{Removing this limit would likely have minimal impact on code performance.}. 
Such a forced step breaks time reversibility, so we count how often it occurs. 
With $M=3$, forced steps occurred in five of the 30 runs: one at $N=400$, $f=2$; two at $N=800$, $f=2$; and three, two, and one at $N=975$ for $f=1$, $2$, and $10$, respectively. 
No forced step occurred for $f=4$ or $f=15$ at any $N$. 
Irreversible steps are therefore rare in this problem.

\begin{figure*}
  \centering
  \includegraphics[width=\textwidth]{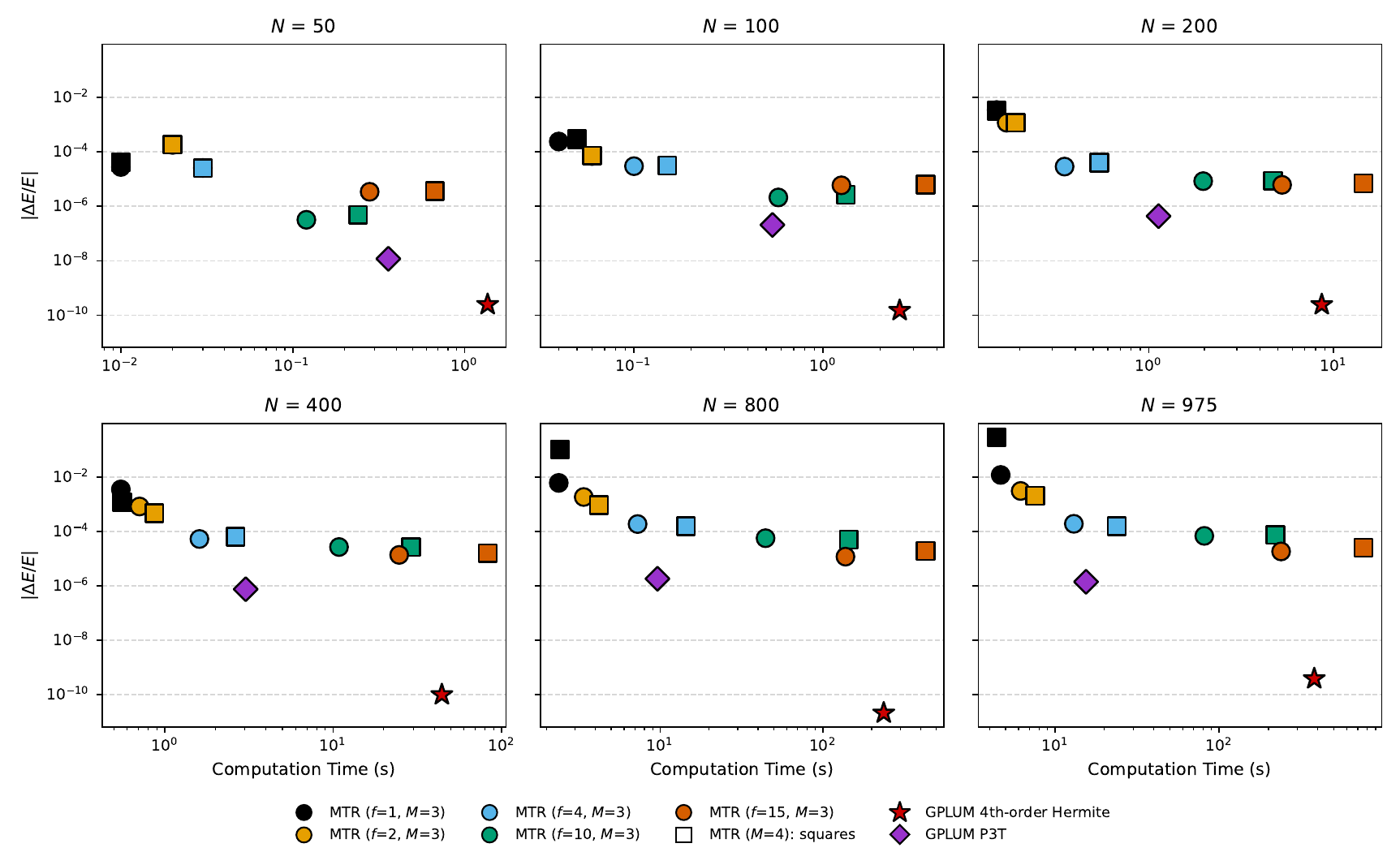}
  \caption{Efficiency diagram for the accretion of the Moon problem: the absolute
  value of the relative energy error at $t=6\pi$ versus wall-clock time, one panel per $N$. Circles are \texttt{MTR} with $M=3$ and squares \texttt{MTR} with $M=4$, coloured by $f$ as in Fig.~\ref{fig:Moon_walltime}. Stars and diamonds are \texttt{GPLUM} in the Hermite and P$^3$T ($\Delta t_{\rm tree}=2^{-7}$) modes, respectively.  }
  \label{fig:Moon_efficiency}
\end{figure*}
%%%%%%%%%%%%%%%%%%%%%%%%%%%%%%%%%%%%%%%%%%%%%%%%%%%%%%%%%%%%%%%%%%%%%%%%%%%%
\subsubsection{Efficiency}
\label{sec:Moon:efficiency}
Fig.~\ref{fig:Moon_efficiency} combines the final energy error and the
computational time of every run in an efficiency diagram.  We plot one panel per $N$. 
Within \texttt{MTR}, increasing $f$ moves the points down (more accuracy) and to the right (more compute time). 
Going from $f=1$ to $f=4$ increases the computational time by a factor of 2--3 at every $N$ (e.g. from 4.7 to 13 s at $N=975$).
The gain in accuracy, as we increase $f$ from $1$ to $4$, depends on $N$. There is a gain of about one order of magnitude at $N=100$, and a gain of one to two orders of magnitude for $N\ge200$.
For $f=10$ and $15$ the absolute value of the error lies between $3\times10^{-7}$ and $7\times10^{-5}$ at every $N$.   The squares show that in most cases $M=4$ gives the same error as $M=3$ at 1--3.5 times the cost, so $M=3$ is the better choice. 
The \texttt{GPLUM} Hermite mode reaches errors of $2\times10^{-11}$--$4\times10^{-10}$, three to six orders of magnitude below the most accurate \texttt{MTR} run at the same $N$, and the P$^3$T mode $10^{-8}$--$2\times10^{-6}$, about one order of magnitude below the most accurate \texttt{MTR} run.  P$^3$T lies below all \texttt{MTR} points at every $N$; it lies to the right of the $M=3$ points with $f\le4$ and, for $N\ge100$, to the left of the $f=10$ and 15 points.

\begin{figure*}
  \centering
 \includegraphics[width=0.85\textwidth]{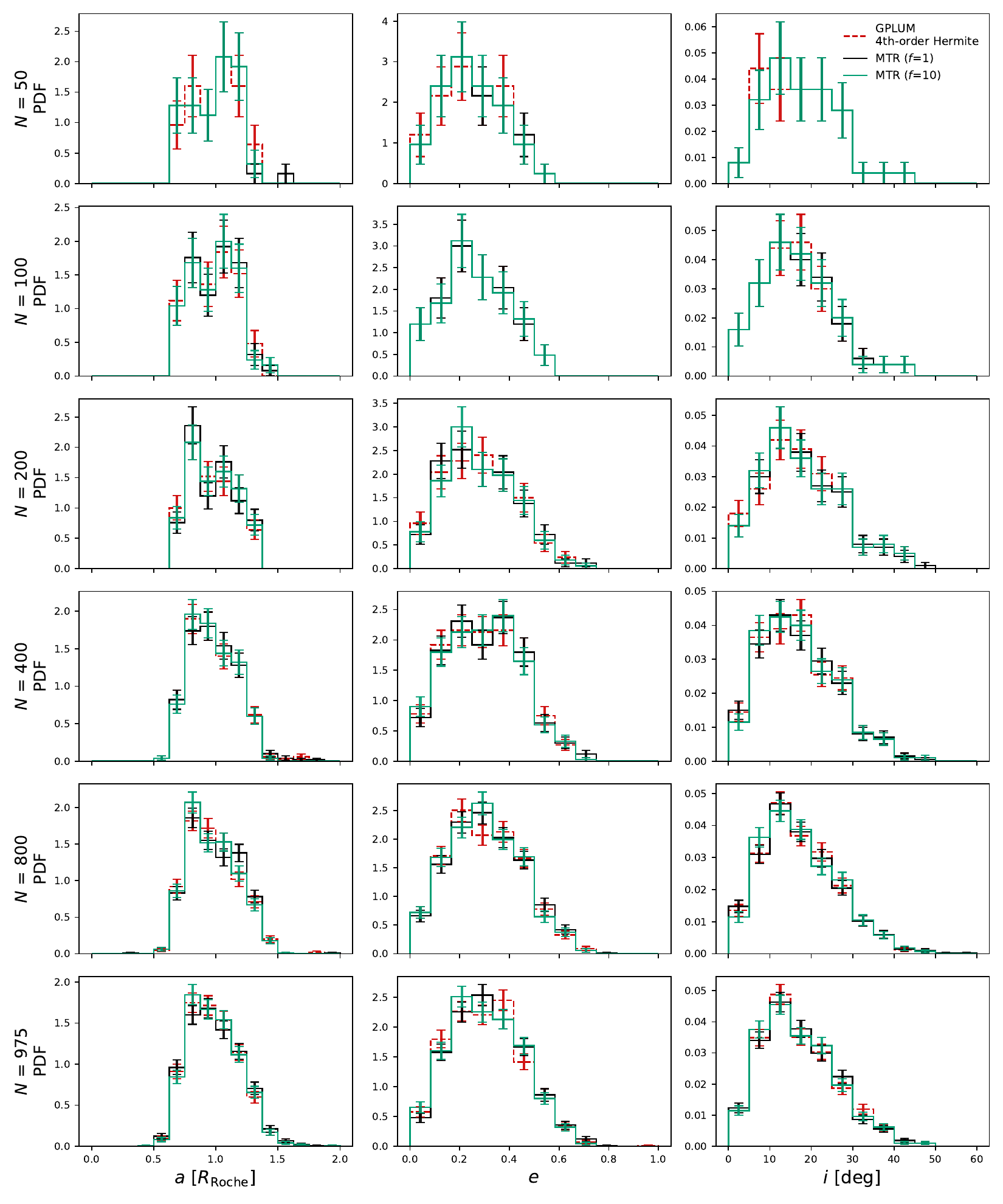}
  \caption{Distributions of semi-major axis (left), eccentricity (middle), and inclination (right) of the disc particles at $t=6\pi$, one row per $N$, for \texttt{GPLUM} in the Hermite mode (red dashed) and for \texttt{MTR} with $M=3$ and $f=1$ (black) and $f=10$ (green). The error bars are the Poisson errors. }
  \label{fig:Moon_pdfs}
\end{figure*}

$f=4$ with $M=3$ appears a good compromise for this problem (see also Fig. \ref{fig:Moon_pdfs}).  Thus, we consider this our default configuration.  We note again that angular momentum is conserved to roundoff error in every run.  This is a notable advantage over \texttt{GPLUM}.

%%%%%%%%%%%%%%%%%%%%%%%%%%%%%%%%%%%%%%%%%%%%%%%%%%%%%%%%%%%%%%%%%%%%%%%%

%\subsubsection{Energy error evolution}
%\label{sec:Moon:energy}
%Fig.~\ref{fig:Moon_energy} shows the time evolution of the energy error for all $M=3$ runs.
%For $f=1$ and 2 the error is a staircase of discrete negative jumps, each
%associated with a single close encounter that is not resolved well enough. 
%At $N=200$ and $f=1$, for example, one event at $t\approx7$ accounts for almost the entire final error of $-3.5\times10^{-3}$.  For $f=4$ the jumps are much smaller once $N\ge100$, and a slow negative drift dominates the error, so that for $N\ge100$ the final error is one to two orders of magnitude smaller than for $f=1$. 

%At $N=800$ the improvement from $f=1$ to $f=15$ is a factor of 500. 

%%%%%%%%%%%%%%%%%%%%%%%%%%%%%%%%%%%%%%%%%%%%%%%%%%%%%%%%%%%%%%%%%%%%%%%%%%%%
\subsubsection{Orbital element distributions}
\label{sec:Moon:PDF}
Fig.~\ref{fig:Moon_pdfs} compares the distributions of $a$, $e$, and $i$ at the end of the runs obtained with \texttt{GPLUM} in the Hermite mode and with \texttt{MTR} at $f=1$ and $10$.
Over three orbits, scattering has spread the semi-major axes beyond the initial range $0.7$--$1.3\,R_{\rm Roche}$. 
The error bars in Fig.~\ref{fig:Moon_pdfs} are the Poisson errors of the bin counts. 
The three runs have very different final energy errors, up to $10^{-2}$ for $f=1$, $10^{-7}$--$10^{-5}$ for $f=10$, and $10^{-11}$--$10^{-10}$ for \texttt{GPLUM} Hermite mode (see Fig.~\ref{fig:Moon_efficiency}). Nevertheless, at every $N$ the three distributions agree within the error bars. 
The large energy error at $f=1$ comes from a few under-resolved close encounters. These encounters change the orbits of a few particles but leave the distributions of the disc as a whole unchanged.
%%%%%%%%%%%%%%%%%%%%%%%%%%%%%%%%%%%%%%%%%%%%%%%%%%%%%%%%%%%%%%
\section{Discussion}
\label{sec:conc}
We have presented and tested a new, (almost) time-reversible gravitational dynamics code, written in \texttt{C}, called \texttt{MTR} (short for Multiple Timestep Reversible).  This code works for planetary systems with a dominant mass (the star).  Gravity admits a large hierarchy of time-scales, as planets may collide, scatter, or evolve slowly on orbital or even secular time-scales.  Hybrid methods provide a blunt mechanism for adjusting to this hierarchy, but \texttt{MTR} attempts an even more efficient approach by assigning every planet pair its own timestep, each of which evolves independently in time.

Careful testing confirms \texttt{MTR} behaves as a time-symmetric method, with excellent energy conservation properties, in addition to machine-precision angular momentum conservation.  We have also explicitly carried out a time-symmetry round-trip test.

\texttt{MTR}'s computational cost scales well with $N$, as $N^2$ roughly.  Compared to \texttt{GPLUM}'s direct $N$-body Hermite mode, for which it is not optimized, \texttt{MTR} achieves a speedup of up to two orders magnitude, while the distribution of orbits still agrees.  Using \texttt{MTR} with $r_{\mathrm{Hill}} = 4 r_{\mathrm{H}}$, where $r_{\mathrm{H}}$ is a mutual Hill radius, and timestep criterion \ref{en:crit1}, the speed is comparable to or up to an order magnitude faster than \texttt{GPLUM} in tree mode.  This is notable because while \texttt{MTR} is a direct $N$-body code with all forces calculated exactly, a tree code approximates long-range forces in order to gain a speedup.  \texttt{MTR} is not intended to, and should not, replace a code like \texttt{GPLUM}.  \texttt{GPLUM} is a P$^3$T code, which scales as $N \log N$: for very large particle numbers, it is a more reasonable choice than a direct $N$-body code like \texttt{MTR}, with $N^2$ scaling.  However, it is feasible to use \texttt{MTR} itself within a tree code for the close neighbour handling part.  \texttt{GPLUM} is also designed for parallel architectures, a major strength.  \texttt{MTR} is not currently a code for parallel machines.  

\texttt{MTR} is provided freely; however, the code is intended as a proof of concept of a many-body time-symmetric code.  There are additional challenges which \texttt{MTR} cannot yet solve; for instance, it fails for orbits with very close pericentre passages to the star.  Extra forces beyond Newtonian point gravity are allowed by the algorithm, but are not currently implemented.  We have also not tested problems with many escaping or ejected particles.  Ejection statistics may be especially sensitive to the accuracy of the close encounter handling \citep{Heggie91}, and it would be interesting to see how \texttt{MTR} handles such situations.  However, in another time-reversible code we have worked on \citetext{\citealp{Lu2024}; Lu et al., in prep}, ejected particle distributions agree well with other codes, and we expect a similar situation with \texttt{MTR}.  \texttt{MTR} can be improved and sped up further in future work by, for example, adapting it to parallel computing.

%%%%%%%%%%%%%%%%%%%%%%%%%%%%%%%%%%%%%%%%%%%%%%%%%%%%%%%%%%%%%%
\section{Acknowledgements}
\texttt{MTR} was originally written in \texttt{MATLAB} by the authors, with some efficiency suggestions from \texttt{ChatGPT}, and was translated to \texttt{C} by \texttt{Claude} AI, with guidance from the authors.  The authors have checked and tested the translation carefully.
{\texttt{Claude} AI was used to assist in debugging the C version of \texttt{MTR} and to check the grammar and clarity of the manuscript text. }  \texttt{Claude} AI was used to assist in creating visual details in figures and for implementing the round-trip reversibility test.  All ideas are due to the authors.

DMH acknowledges support from the National Science and Technology Council in Taiwan through grant 114-2112-M-003 -020-MY3, and from the Taiwanese Ministry of Education Yushan Fellowship, MOE-115-YSFMS-0004-002-P1.

%%%%%%%%%%%%%%%%%%%%%%%%%%%%%%%%%%%%%%%%%%%%%%%%%%%%%%%%%%%%%%
\section{Data Availability}
The \texttt{MTR} code is freely available along with a set of test problems and usage instructions \footnote{https://github.com/dmhernan/Multiple-Timestep-Reversible}.

%%%%%%%%%%%%%%%%%%%%%%%%%%%%%%%%%%%%%%%%%%%%%%%%%%

%%%%%%%%%%%%%%%%%%%%%%%%%%%%%
\bibliographystyle{mnras}
\bibliography{paper}
%%%%%%%%%%%%%%%%%%%%%%%%%%%%%%%%%%%%%%%%%%%%%%
%\clearpage
\appendix
\onecolumn

%%%%%%%%%%%%%%%%%%%%%%%%%%%%%%%%%%%%%%%%%%%%%%
\section{Table of Symbols}
\label{app:symbols}

\noindent
\begin{minipage}{\textwidth}
\makeatletter\def\@captype{table}\makeatother
\centering
\renewcommand{\arraystretch}{1.3}
\setlength{\tabcolsep}{8pt}
\small
\caption{Important symbols and functions used in this work, along with their reference and description.}
\label{tab:symbols}
\begin{tabular}{p{2.8cm} p{9.5cm} p{2.8cm}}
\hline
Symbol & Description & Reference \\
\hline
$\v{Q}_{i\ne 0}$    & Planet $i$ heliocentric position & Section~\ref{sec:math} \\
$\v{P}_{i\ne 0}$    & Planet $i$ barycentric momentum & Section~\ref{sec:math} \\
$\v{Q}_{ij}$        & $\v{Q}_i - \v{Q}_j$ & Section~\ref{sec:math} \\
$\v{z}$             & $(\v{Q},\v{P})$ & Section~\ref{sec:math} \\
$m_0$               & Mass of star & Section~\ref{sec:math} \\
$m_{i \ne 0}$       & Mass of planet $i$ & Section~\ref{sec:math} \\
$H$                 & $H = H_\mathrm{Kep} + H_\mathrm{Sun} + V$ & Section~\ref{sec:math} \\
$H_{\mathrm{Kep},i}$ & Planet $i$ Kepler Hamiltonian & Equation~\eqref{eq:Hdef} \\
$H_\mathrm{Sun}$    & Kinetic energy of star & Equation~\eqref{eq:Hdef} \\
$V$                 & Planet--planet potential & Equation~\eqref{eq:Hdef} \\
$V_{ij}$            & Potential between planet $i$ and planet $j$ & Equation~\eqref{eq:opdef} \\
$h_0$               & Global timestep & Equation~\eqref{eq:psi2} \\
$h_i$               & $h_0 / M^i$ & Section~\ref{sec:math} \\
$s_i$               & $i$th substep & Equation~(\ref{eq:adap}) \\
$M$                 & Timestep ratio between consecutive timestep levels & Section~\ref{sec:math} \\
$R$                 & Ratio between consecutive cutoff radii $r_i$ & Section~\ref{sec:math} \\
$r_{i \ge 1}$       & $r_0/R^{i}$ & Equation~(\ref{eq:rshell}) \\
$\sub{r}{Hill}$     & Hill radius or characteristic scale & Section~\ref{sec:tcrit} \\
$\phi(s_i)$         & Kick--Kepler--kick substep map at substep $s_i$ & Equation~\eqref{eq:phi} \\
$\hat{A}_i$         & Kick operator at level $i$ & Equation~(\ref{eq:opdef}) \\
$\hat{B}_i$         & Kepler solver at level $i$ & Equation~(\ref{eq:opdef}) \\
$S_i$               & Set of pairs to be kicked at level $i$ & Equation~(\ref{eq:opdef}) \\
$S_i^\prime$        & Planet indices at level $i$ getting a Kepler solver & Equation~(\ref{eq:opdef}) \\
$\hat{D}_i$         & Map in \texttt{MTR} & Equation~(\ref{eq:Cmap}) \\
$\hat{D}_0$         & \texttt{MTR}, the main result of this paper & Equation~(\ref{eq:Cmap}) \\
\texttt{levc}       & Matrix with the timesteps of each pair & Section~\ref{sec:codelog} \\
\texttt{levcp}      & Timesteps of each pair at the end of the substep & Section~\ref{sec:codelog} \\
\texttt{F}          & $\texttt{levcp} - \texttt{levc}$ & Section~\ref{sec:codelog} \\
\texttt{arru}       & Array of pairs getting treated at levels $\ge l$, where $l$ is the current level & Section~\ref{sec:codelog} \\
\texttt{MTR}        & Multiple Timestep Reversible code, the main result of this work & Section~\ref{sec:math} \\
\hline
\end{tabular}
\end{minipage}

%%%%%%%%%%%%%%%%%%%%%%%%%%%%%%%%%%%%%%%%%%%%%%%%%%%%%%%%%%%%%%%%
\end{document}